\documentclass[twocolumn,a4paper,11pt]{article}

\usepackage{amsmath}
\usepackage[psamsfonts]{amssymb}
\usepackage[dvips]{graphicx}
\usepackage{color}

\begin{document} 

\begin{titlepage}

\baselineskip 10pt
\hrule 
\vskip 5pt
\leftline{}
\leftline{
          \hfill   \small \hbox{\bf CHIBA-EP-147}}
\leftline{\hfill   \small \hbox{hep-th/0407210}}
\leftline{\hfill   \small \hbox{April 2005}}
\vskip 5pt
\baselineskip 14pt
\hrule 
\vskip 1.0cm
\centerline{\Large\bf 
Understanding the U(1) problem 
} 
\vskip 0.5cm
\centerline{\Large\bf  
through dyon configuration   
}
\vskip 0.5cm
\centerline{\Large\bf  
in the Abelian projection 
}
\vskip 0.3cm
\centerline{\large\bf  
}

\vskip 0.5cm

\centerline{{\bf 
Seikou Kato$^{\dagger,{1}}$ and 
Kei-Ichi Kondo$^{\ddagger,{2}}$ 
}}  
\vskip 0.5cm
\centerline{\it
${}^{\dagger}$ 
Takamatsu National College of Technology,
Takamatsu City 761-8058, Japan
}
\centerline{\it
${}^{\ddagger}$Department of Physics, Faculty of Science, 
Chiba University, Chiba 263-8522, Japan
}
\vskip 1cm

\begin{abstract}
We show that the magnetic monopole promoted to the dyon due to the vacuum angle $\theta$ resolves the U(1) problem
in the sense that the dyon obtained in this way gives a dominant contribution to the topological susceptibility. 
For this purpose, we derive an Abelian-projected effective gauge theory written in terms of Abelian degrees of freedom, which is obtained by integrating out all the off-diagonal degrees of freedom involved in the SU(2) Yang-Mills theory with the vacuum angle $\theta$.
We  evaluate the topological susceptibility by estimating the classical part of the effective dyon action obtained by performing the duality transformation. 
The obtained result is consistent with the Veneziano--Witten formula. 

\end{abstract}


\hrule  
\vskip 0.1cm
${}^1$ 
  E-mail:  {\tt kato@takamatsu-nct.ac.jp}

${}^2$ 
  E-mail:  {\tt kondok@faculty.chiba-u.jp}

\par 
\par\noindent


\vskip 0.5cm

\newpage
\pagenumbering{roman}
\tableofcontents




\end{titlepage}


\pagenumbering{arabic}

\baselineskip 14pt
\section{Introduction}

The elementary particles constituting hadrons such as baryons and mesons are called quarks, which are combined by gluons as gauge particles. 
Nowadays, the very fundamental theory describing quarks and gluons is believed to be quantum chromodynamics (QCD) which is a non-Abelian gauge theory or the Yang-Mills theory with color gauge group SU(3). 
This is because QCD is the unique theory which well describes the asymptotic behavior of hadrons in high energy region reflecting the asymptotic freedom using the perturbation theory  and automatically satisfies a number of conservation laws characteristic in the strong interactions.

On the other hand,  the perturbation theory is powerless to study the hadronic phenomena in the low-energy region where the coupling constant becomes large.
For instance, it is difficult to calculate the proton mass directly from QCD.  
Moreover, quarks have never been observed in the isolated form, which is called the quark confinement problem. 

There is another problem called the U(1) problem or $\eta$ meson problem \cite{Weinberg75}. 
The $\eta'$ meson is regarded as a Nambu--Goldstone (NG) boson (pseudo-scalar) associated with the spontaneous breaking of chiral symmetry to the flavor symmetry, $U(N_f)_L \times U(N_f)_R \to U(N_f)_V$, 
caused by the flavor-independent quark-antiquark pair condensations
$\langle \bar{u}u \rangle 
=\langle \bar{d}d \rangle 
=\langle \bar{s}s \rangle$. 
However, $\eta'$ is very heavy compared with the other eight NG bosons,
$\pi^0, \pi^\mp, K^\pm, K^0, \bar{K}^0, \eta$, for $N_f=3$. 
In fact, the mass of $\eta'$ meson is about $958\rm{\rm{MeV}}$ which is about 10 times larger than the mass   $135\rm{MeV}$ of $\pi$ meson as one of the NG bosons. 
Moreover, there are inconsistencies between the theoretical prediction and the experimental data for the decay of the $\eta$ meson, e.g.,
$\eta \rightarrow \pi^+ \pi^- \pi^0$.
These problems have a common origin, i.e.,  the singularity of the color-flavor singlet axial-vector current (so-called the U$_A$(1) current).  

We know other problems to be resolved, such as strong CP violation and  chiral symmetry breaking in the strong interactions. 
In order to solve these problems, we need to develop the non-perturbative methods without relying on the perturbation theory. 
These non-perturbative phenomena are believed to be well understood in the unified way by considering the topologically nontrivial configurations of the gluon field.

The global U$_A$(1) symmetry is broken at the quantum level, since the U$_A$(1) current has the triangle anomaly in the quantum theory.  
In fact, 't Hooft \cite{tHooft76} pointed out that topologically nontrivial configurations such as instantons give the nonzero anomaly 
and suggested that instantons are the relevant topological objects related to the resolution of the U(1) problem\cite{instantonGroup}. 
However, it was not clear how to  compute the $\eta'$ mass. 
Moreover, it was pointed out that the Ward-Takahashi identity for the U$_A$(1) current with the anomalous term contradicts with the quark--antiquark condensation in the instanton $\theta$ vacuum \cite{Crewther77}.

There is another route initiated by Witten \cite{Witten:1979ey1} and Veneziano \cite{Veneziano79} for solving the U(1) problem within the framework of the large $N_c$ (color) expansion.  They have derived the relation called the Witten--Veneziano formula which enables us to estimate the $\eta'$ mass through the topological susceptibility. 
Along this line,  a lot of progress have been made by subsequent works \cite{U1}.  
Nowadays, it is recognized as a solution of the U(1) problem.

In this paper, we argue that the U(1) problem is understood through the
dyon configuration. 
A strategy for solving the U(1) problem along this line has already been discussed by Ezawa and Iwazaki~\cite{EI82} based on the idea of the Abelian projection proposed by 't~Hooft~\cite{tHooft81}. 
However, they assumed   in their analyses the {\it Abelian dominance} from the beginning and used  an Abelian-projected effective theory which is conjectured to be derived from the Yang-Mills theory in the long distance.  
In contrast, in this paper, {\it we derive the Abelian-projected effective theory} based on the functional integration of the off-diagonal degrees of freedom from the Yang-Mills theory with the $\theta$ angle. 

This paper is organized as follows. 
In section 2, we adopt the BRST formulation to quantize the Yang-Mills theory with the $\theta$ angle where we restrict our consideration to the gauge group SU(2) for simplicity.
We exploit the Abelian projection idea \cite{tHooft81} and integrate out all the off-diagonal components of gluons.  Then we obtain an   effective theory written in terms of the diagonal gluons alone, which we call the Abelian Projected Effective Gauge Theory (APEGT) \cite{Kondo97} with the $\theta$ angle.  
The Abelian projection here does not mean that the off-diagonal gluons are simply neglected to obtain the APEGT for studying the low-energy physics. In fact,  the off-diagonal gluons influence the wavefunction renormalization and the running effective coupling constant in the resulting APEGT to be consistent with the asymptotic freedom in the original Yang-Mills theory \cite{Kondo97}.

In section 3, we rewrite the APEGT with $\theta$-term into an effective theory written in terms of the dyon degrees of freedom alone.  
The obtained dyon action has a beautiful form suggesting the existence of the duality in the effective Abelian gauge theory. 
Here the dyon implies a topological soliton having both  electric and magnetic charges where  
the electric charge of dyon is proportional to the $\theta$ angle.

In section 4, we evaluate the topological susceptibility from the effective dyon action.  We show that the U(1) problem is solved by using the effective dyon action obtained in this way, if it is combined with the Witten--Veneziano formula.

In the final section, we discuss the relationship between the dyon and the instanton from the viewpoint of understanding the U(1) problem.

\renewcommand{\thefootnote}{\fnsymbol{footnote}}
\newcommand{\qed}{\hbox{\rule[-2pt]{3pt}{6pt}}}
\newcommand{\eq}[1]{(\ref{#1})}
\newcommand{\la}[1]{\label{eqn:#1}}
\newcommand{\beq}{\begin{align*}}
\newcommand{\eeq}{\end{align*}}
\newcommand{\nn}{\nonumber}
\newcommand{\intpi}{\int\limits_{-\pi}^{+\pi}}
\newcommand{\intinf}{\int\limits_{-\infty}^{+\infty}}
\newcommand{\dd}{\partial}
\newcommand{\dD}{{\cal D}}
\newcommand{\dS}{{\cal S}}
\newcommand{\dZ}{{\cal Z}}
\newcommand{\sbra}[1] { \left( #1 \right)}     
\newcommand{\mbra}[1] { \left\{ #1 \right\}}     
\newcommand{\lbra}[1] { \left[ #1 \right]}     
\newcommand{\zbra}[1] { \left| #1 \right|}     
\newcommand{\bra}[1] { \left\| #1 \right\|}     
\newcommand{\kbra}[1] { \left< #1 \right>}     
\newcommand{\vct}[3]{ \left(
                        \begin{array}{c}
                           #1 \\ #2 \\ #3 
                        \end{array}
                       \right) }               
\newcommand{\vect}[2]{ \left(
                        \begin{array}{c}
                           #1 \\ #2 
                        \end{array}
                       \right) }               
\newcommand{\mtx}[4]{ \left(
                      \begin{array}{cc}
                          #1 & #2 \\
                          #3 & #4 
                      \end{array}
                       \right) }       
\newcommand{\bh}{{}^{-}\!\!\!\!h}
\newcommand{\ket}[1]{|#1>}
\newcommand{\brav}[1]{<#1|}
\newcommand{\braket}[2]{<#1|#2>}
\renewcommand{\topfraction}{1}  
\renewcommand{\textfraction}{0} 
\renewcommand{\floatpagefraction}{1}  

\section{APEGT with $\theta$ term}

We extend the method of \cite{Kondo97} to the Yang-Mills theory in the presence of the vacuum angle $\theta$. 

\subsection{Definitions}

In this paper, we restrict our consideration to the gauge group $G=SU(2)$. 
We write the SU(2) gluon field ${\cal A}_{\mu}$ as
\begin{align} 
{\cal A}_{\mu}(x) = \sum_{A=1}^{3}{\cal A}_{\mu}^A(x)T^A ,
\label{model-1d0}
\end{align}
and the field strength ${\cal F}_{\mu\nu}$ as
\begin{align} 
{\cal F}_{\mu\nu}(x) &= \sum_{A=1}^{3} {\cal F}^{A}_{\mu\nu}(x)T^{A} 
\nonumber\\&
= \dd_{\mu} {\cal A}_{\nu}(x) - \dd_{\nu} {\cal A}_{\mu}(x) \nn\\
& -ig[{\cal A}_{\mu}(x),{\cal A}_{\nu}(x)] ,
\label{model-1d1}
\end{align}
where $T^A(A=1,2,3)$ is the generator of the Lie algebra of the gauge group $SU(2)$.  The Hodge dual $\tilde{\cal F}_{\mu\nu}$ of ${\cal F}_{\mu\nu}$  is defined by 
\begin{align} 
\tilde{\cal F}_{\mu\nu}(x) &\equiv 
\frac{1}{2}\epsilon_{\mu\nu\alpha\beta}{\cal F}_{\alpha\beta}(x) .
\label{model-1d1-1}
\end{align}

We adopt the Yang-Mills (YM) action ${\cal S}_{YM}[{\cal A}]$ with the $\theta$ term ${\cal S}_{\theta}[{\cal A}]$:
\begin{subequations}
\begin{align} 
{\cal S}_{YM\theta}[{\cal A}] &=
{\cal S}_{YM}[{\cal A}]+{\cal S}_{\theta}[{\cal A}] ,
\label{model-1a}\\
{\cal S}_{YM}[{\cal A}] &= 
-\frac{1}{2g^2}\int_x {\rm tr}({\cal F}_{\mu\nu}{\cal F}^{\mu\nu}) ,
\label{model-1b}\\
{\cal S}_{\theta}[{\cal A}] &= \frac{\theta}{16\pi^2}
\int_x {\rm tr}({\cal F}_{\mu\nu}\tilde{\cal F}^{\mu\nu}) ,
\label{model-1c}
\end{align}
\end{subequations}
where we have introduced the notation, 
$\int_x\equiv \int d^4 x$.

The topological term ${\cal S}_{\theta}[{\cal A}]$
can be cast into the total derivative and is neglected in the perturbation theory.  For the instanton solution with the nontrivial winding number $Q\not=0$, however, it gives a non-trivial value, 
${\cal S}_{\theta}[{\cal A}]=\theta Q \ne 0$.
Therefore, the topological term is expected to give a non-trivial contribution in the non-perturbative phenomena in which the topological configuration such as instanton plays the important role. 

Here, we decompose ${\cal A}_{\mu}$ into the diagonal U(1)  and the off-diagonal SU(2)/U(1) parts as
\begin{align} 
{\cal A}_{\mu}(x)=a_{\mu}(x)T^3+ A_\mu(x) , 
\nonumber\\
   A_\mu(x):=\sum_{a=1}^{2}A_{\mu}^a(x)T^a ,
\label{model-1d}
\end{align}
where the index $a=1,2$ denotes the off-diagonal part. $a_{\mu}(x)$ and 
$A_{\mu}^a(x)$ are diagonal, off-diagonal gluon field, respectively.
Accordingly, the field strength ${\cal F}_{\mu\nu}$ is decomposed as 
\begin{subequations}
\begin{align} 
{\cal F}_{\mu\nu} &= 
[f_{\mu\nu}(x)+{\cal C}_{\mu\nu}(x)]T^3 \nn\\
&+{\cal S}_{\mu\nu}^a(x)T^a, \\
f_{\mu\nu}(x) &\equiv \dd_{\mu}a_{\nu}(x)-\dd_{\nu}a_{\mu}(x), \\
{\cal S}_{\mu\nu}^a(x) &\equiv D_{\mu}[a]^{ab}A_{\nu}^b(x)
-D_{\nu}[a]^{ab}A_{\mu}^b(x),\\
{\cal C}_{\mu\nu}(x)T^3 &\equiv -i[A_{\mu}(x),A_{\nu}(x)] ,
\label{model-2}
\end{align}
where the covariant derivative $D_{\mu}[a]$ is defined by
\begin{align} 
D_{\mu}[a]&=\dd_{\mu}+i[a_{\mu}T^3,\cdot] , \\
D_{\mu}[a]^{ab}&=\dd_{\mu}\delta^{ab}-\epsilon^{ab3}a_{\mu} .
\label{model-3}
\end{align}
\end{subequations}
Then the action is decomposed as
\begin{subequations}
\begin{align} 
{\cal S}_{YM}[{\cal A}] &=
-\frac{1}{4g^2}\int_x [(f_{\mu\nu}+{\cal C}_{\mu\nu})^2 \nn\\
&+({\cal S}_{\mu\nu}^a)^2], 
\label{model-4a}\\
{\cal S}_{\theta}[{\cal A}] &= 
\frac{\theta}{32\pi^2}
\int_x [(f_{\mu\nu}+{\cal C}_{\mu\nu})(\tilde{f}^{\mu\nu}
+\tilde{{\cal C}}^{\mu\nu}) \nn\\
&+{\cal S}_{\mu\nu}^a\tilde{{\cal S}}^{\mu\nu}{}^a] ,
\label{model-4b}
\end{align}
where 
 $({\cal S}_{\mu\nu}^a)^2$
and 
${\cal S}_{\mu\nu}^a\tilde{{\cal S}}^{\mu\nu}{}^a$
are 
\begin{align} 
({\cal S}_{\mu\nu}^a)^2 &= -2A^{\mu}{}^aW_{\mu\nu}^{ab}A^{\nu}{}^b
+2\dd^{\mu}(A^{\nu}{}^a {\cal S}_{\mu\nu}^a), 
\label{su2u1-2}\\
W_{\mu\nu}^{ab} &= (D^{\rho}[a]D_{\rho}[a])^{ab}\delta_{\mu\nu}
-\epsilon^{ab3}f_{\mu\nu} \nn \\
& -D_{\mu}[a]^{ac}D_{\nu}[a]^{cb}, 
\label{su2u1-3}\\
{\cal S}_{\mu\nu}^a\tilde{{\cal S}}^{\mu\nu}{}^a &=
-2A^{\mu}{}^a\tilde{W}_{\mu\nu}^{ab}A^{\nu}{}^b  \nn \\
&+2\dd^{\mu}(A^{\nu}{}^a\cdot 
\frac{1}{2}\epsilon_{\mu\nu\rho\sigma}{\cal S}^{\sigma\rho}),
\label{su2u1-4}\\
\tilde{W}_{\mu\nu}^{ab} &= -\epsilon_{\mu\nu\alpha\beta}
D_{\alpha}[a]^{ac}D_{\beta}[a]^{cb} \nn\\
&=\epsilon^{ab3}\tilde{f}_{\mu\nu}.
\label{su2u1-5}
\end{align}
\end{subequations}
In eq.(\ref{su2u1-5}), we have used 
\begin{align} 
[D_{\mu}[a]^{ac},D_{\nu}[a]^{cb}] = -\epsilon^{ab3}f_{\mu\nu}
\label{su2u1-6}
\end{align}
In what follows, the surface terms, i.e., the second terms in (\ref{su2u1-2}) and (\ref{su2u1-4}), are neglected, since it is known that the off-diagonal gluons become massive once the MA gauge fixing is adopted\cite{AS99}.

Thus, the total action ${\cal S}_{YM\theta}[{\cal A}]$
is decomposed as
\begin{align} 
& {\cal S}_{YM\theta}[{\cal A}] =
\int_x \left\{
-\frac{1}{4g^2}f_{\mu\nu}f^{\mu\nu}
+\frac{\theta}{32\pi^2}f_{\mu\nu}\tilde{f}^{\mu\nu} 
\right. \nn\\
&\quad\quad \left. 
+f_{\mu\nu}\lbra{-\frac{1}{2g^2}{\cal C}^{\mu\nu}+
\frac{\theta}{16\pi^2}\tilde{{\cal C}}^{\mu\nu}} 
\right. \nonumber \\
&\quad\quad \left.
-\frac{1}{4g^2}{\cal C}_{\mu\nu}{\cal C}^{\mu\nu}
+\frac{\theta}{32\pi^2}{\cal C}_{\mu\nu}\tilde{{\cal C}}^{\mu\nu}
\right. \nn\\
&\quad\quad \left. 
+\frac{1}{2g^2}A^{\mu}{}^a
\lbra{W_{\mu\nu}^{ab}
-\frac{g^2\theta}{8\pi^2}\epsilon^{ab3}\tilde{f}_{\mu\nu}}
A^{\nu}{}^b
\right\}.
\label{model-5}
\end{align}
In order to integrate out the off-diagonal gluon field $A^a_{\mu}$, 
we replace  the terms quartic in $A^a_{\mu}$ in (\ref{model-5}),
\begin{align} 
\int_x\mbra{-\frac{1}{4g^2}{\cal C}_{\mu\nu}{\cal C}^{\mu\nu}
+\frac{\theta}{32\pi^2}{\cal C}_{\mu\nu}\tilde{{\cal C}}^{\mu\nu}} ,
\label{model-6}
\end{align}
by the equivalent form quadratic in $A^a_{\mu}$,
\begin{align} 
\int_x\mbra{-\frac{1}{4}g^2B_{\mu\nu}B^{\mu\nu}
+\frac{1}{2}B_{\mu\nu}
\lbra{c_0{\cal C}^{\mu\nu}+c_1\tilde{{\cal C}}^{\mu\nu}}} ,
\label{model-7}
\end{align}
with appropriate constants, $c_0$ and $c_1$, to be specified shortly.%
\footnote{ 
This procedure corresponds to introducing the auxiliary (antisymmetric tensor) field $B_{\mu\nu}$ 
according to 
\begin{align} 
B_{\mu\nu}=g^{-2}(c_0{\cal C}_{\mu\nu}
+c_1\tilde{{\cal C}}_{\mu\nu}).
\label{model-7b}
\end{align}
}
In the Minkowski spacetime, by paying attention to the relationship for the double Hodge-dual operations, 
\begin{align} 
\tilde{{\cal C}}_{\mu\nu}\tilde{{\cal C}}^{\mu\nu}
=\frac{1}{4}\epsilon_{\mu\nu\rho\sigma}\epsilon^{\mu\nu\alpha\beta}
{\cal C}^{\rho\sigma}{\cal C}_{\alpha\beta}
=-{\cal C}_{\mu\nu}{\cal C}^{\mu\nu} ,
\label{model-8}
\end{align}
the Gaussian integration over the $B_{\mu\nu}$ field 
in (\ref{model-7}) is performed to give 
\begin{align} 
& -\frac{1}{4}g^2B_{\mu\nu}B^{\mu\nu}
+\frac{1}{2}B_{\mu\nu}
\lbra{c_0{\cal C}^{\mu\nu}+c_1\tilde{{\cal C}}^{\mu\nu}} \nn\\
&\to 
\frac{1}{4g^2}\mbra{
(c_0^2-c_1^2){\cal C}_{\mu\nu}{\cal C}^{\mu\nu} 
+2c_0c_1{\cal C}_{\mu\nu}\tilde{{\cal C}}^{\mu\nu}} .
\label{model-9}
\end{align}
Therefore, in order for (\ref{model-6}) and (\ref{model-7}) to be equivalent, two coefficients, $c_0$ and $c_1$, must satisfy the relationships,%
\footnote{
In the Euclidean space, it should be remarked that the first equation in (\ref{model-10})
has the different form, $c_1^2+c_0^2=-1$,  
due to 
$\tilde{{\cal C}}_{\mu\nu}\tilde{{\cal C}}_{\mu\nu}
={\cal C}_{\mu\nu}{\cal C}_{\mu\nu}$. 
} 
\begin{align} 
c_1^2-c_0^2=1, \quad c_0c_1=\frac{g^2\theta}{16\pi^2} .
\label{model-10}
\end{align}
The solution of (\ref{model-10}) is 
\begin{align} 
c_0^2=\frac{1}{2}\sbra{-1\pm \frac{g^2}{4\pi}|\tau|}, \quad
c_1^2=\frac{1}{2}\sbra{1\pm \frac{g^2}{4\pi}|\tau|} , 
\label{model-11}
\end{align}
where $\tau$ is the complex coupling constant defined by 
\begin{align} 
\tau\equiv \frac{\theta}{2\pi}+i\frac{4\pi}{g^2},\quad
|\tau|=\sqrt{\sbra{\frac{\theta}{2\pi}}^2
+\sbra{\frac{4\pi}{g^2}}^2} ,
\label{model-12}
\end{align}
which is known to play the very important role especially in the supersymmetric Yang-Mills theory, see e.g., \cite{Harvey96}.

In what follows, we adopt 
\begin{align} 
c_0 &= \sqrt{\frac{1}{2}
\sbra{-1+\sqrt{1+\sbra{\frac{g^2\theta}{8\pi^2}}^2}}}
,
\nonumber\\ 
c_1 &= \sqrt{\frac{1}{2}
\sbra{1+\sqrt{1+\sbra{\frac{g^2\theta}{8\pi^2}}^2}}} .
\label{model-13}
\end{align}
For  $g^2\theta << 1$, we see the coefficients behave as
\begin{align} 
c_0 \simeq \frac{1}{2}\cdot \frac{g^2\theta}{8\pi^2}
,\quad 
c_1  \simeq 1+ \frac{1}{8}\sbra{\frac{g^2\theta}{8\pi^2}}^2 .
\label{model-14}
\end{align}

Replacing (\ref{model-6}) with (\ref{model-7}) and using 
\begin{align} 
& f_{\mu\nu}\lbra{-\frac{1}{2g^2}{\cal C}^{\mu\nu}+
\frac{\theta}{16\pi^2}\tilde{{\cal C}}^{\mu\nu}} \nn\\
& \quad\quad +\frac{1}{2}B_{\mu\nu}
\lbra{c_0{\cal C}^{\mu\nu}+c_1\tilde{{\cal C}}^{\mu\nu}} 
\nonumber \\
& =  -\frac{1}{2g^2}A^{\mu}{}^a \left\{
\epsilon^{ab3}f_{\mu\nu}
-\frac{g^2\theta}{8\pi^2}\epsilon^{ab3}\tilde{f}_{\mu\nu}
\right. \nn 
\\
& \quad\quad \left.
-g^2c_0\epsilon^{ab3}B_{\mu\nu}
-g^2c_1\epsilon^{ab3}\tilde{B}_{\mu\nu}
\right\}A^{\nu}{}^b ,
\label{model-15}
\end{align}
the total action reads
\begin{align}
&{\cal S}_{YM\theta}[{\cal A}] \nn\\
&= \int_x
\left\{
-\frac{1}{4g^2}f_{\mu\nu}f^{\mu\nu}
+\frac{\theta}{32\pi^2}f_{\mu\nu}\tilde{f}^{\mu\nu}
\right. \nn\\
& \left.
-\frac{1}{4}g^2B_{\mu\nu}B^{\mu\nu}
+\frac{1}{2g^2}A^{\mu}{}^aQ_{\mu\nu}^{ab}A^{\nu}{}^b 
\right\}  ,
\label{model-16a}
\end{align}
where we have defined 
\begin{align}
Q_{\mu\nu}^{ab} &:=  (D_{\rho}[a]D_{\rho}[a])^{ab}\delta_{\mu\nu}
-2\epsilon^{ab3}f_{\mu\nu} \nn \\
& +g^2c_1\epsilon^{ab3}\tilde{B}_{\mu\nu}
+g^2c_0\epsilon^{ab3}B_{\mu\nu} \nn\\
&-D_{\mu}[a]^{ac}D_{\nu}[a]^{cb}.
\label{model-16b}
\end{align}

\subsection{Gauge fixing}

We adopt the gauge fixing (GF) condition for the off-diagonal part:
\begin{align} 
F^{\pm}[A,a] \equiv (\dd^{\mu}\pm i\xi a^{\mu})A_{\mu}^{\pm}=0 ,
\label{gf1}
\end{align}
where we have used the $(\pm,3)$ basis,
$
{\cal O}^{\pm} \equiv ({\cal O}^1 \pm i {\cal O}^2)/\sqrt{2} .
$
\ Here the gauge parameter $\xi=0$ corresponds to the Lorentz gauge and $\xi=1$ to (the differential form of) the maximal abelian gauge (MAG). 
At this stage, we keep the residual U(1) gauge invariance without fixing it. 

In the BRST quantization, the GF condition (\ref{gf1}) amount to adding the following GF term and the Faddeev--Popov (FP) term \cite{Kondo97},
\begin{align} 
{\cal L}_{GF+FP} &= \phi^aF^a[A,a]+\frac{\alpha}{2}(\phi^a)^2 
\nn\\
&+i\bar{c}^aD^{\mu ab}[a]^{\xi}D_{\mu}^{bc}[a]c^c 
\nonumber \\
&-i\xi \bar{c}^a[A_{\mu}^aA^{\mu b}-A_{\mu}^cA^{\mu c}\delta^{ab}]c^b ,
\label{gf3}
\end{align}
where
\begin{align} 
F^a[A,a]&=(\dd^{\mu}\delta^{ab}-\xi\epsilon^{ab3}a^{\mu})A_{\mu}^b
\nn\\
&= D^{\mu ab}[a]^{\xi}A_{\mu}^b .
\label{gf4}
\end{align}
Thus the total Lagrangian is obtained by adding  (\ref{model-16a}) to (\ref{gf3}),
\begin{align} 
{\cal L} = {\cal L}_{apYM}[{\cal A},\theta] + {\cal L}_{GF+FP}.
\label{gf5}
\end{align}

\subsection{Integration over all SU(2)/U(1) components}

Now we integrate out the off-diagonal fields, $\phi^a$, $A_{\mu}^a$, $c^a$, $\bar{c}^a$ belonging to SU(2)/U(1) and obtain the Abelian-projected effective gauge theory (APEGT) written in terms of the diagonal fields, $a_{\mu}$ and $B_{\mu\nu}$.

\underline{Integrating the Lagrange multiplier field $\phi^a$}\\
 For $\alpha\ne 0$, the Gaussian integration over $\phi^a$ can be done with ease as 
\begin{align} 
& \phi^aF^a[A,a]+\frac{\alpha}{2}(\phi^a)^2
\nn \\
& \quad\quad \to -\frac{1}{2\alpha}(F^a[A,a])^2
 .
\label{su2u1-1}
\end{align}

\underline{Integrating the off-diagonal gluon field $A_{\mu}^a$}\\
For the gauge parameter $\alpha=1$, the total action in the MAG ($\xi=1$) reads%
\footnote{
We can introduce the source term ${\cal A}_{\mu}J^\mu$ with the source $J^\mu$ to obtain the generating functional as in \cite{Kondo97}.
} 
\begin{align} 
S_{YM\theta} &= S_{YM\theta}[a,A,B,c,\bar{c};\theta] \nn \\
&= S_1[a,B;\theta]+S_2[a,c,\bar{c}] \nn \\
&+S_3[a,A,B,c,\bar{c};\theta],
\\
S_1 &= \int_x [-\frac{1}{4g^2}f_{\mu\nu}f^{\mu\nu}
+\cdot \frac{\theta}{32\pi^2}f_{\mu\nu}\tilde{f}^{\mu\nu} 
\nn \\
& -\frac{1}{4}g^2B_{\mu\nu}B^{\mu\nu}], 
\label{su2u1-7}\\
S_2 &= \int_x i\bar{c}^aD^{\mu ac}[a]D_{\mu}^{cb}[a]c^b,
\label{su2u1-8}\\
S_3 &= \int_x \frac{1}{2g^2}A^{\mu}{}^a Q_{\mu\nu}^{ab}A^{\nu}{}^b,
\label{su2u1-9}
\\
Q_{\mu\nu}^{ab} &\equiv  (D_{\rho}[a]D_{\rho}[a])^{ab}\delta_{\mu\nu}
-2\epsilon^{ab3}f_{\mu\nu}
\nn\\
&+g^2c_1\epsilon^{ab3}\tilde{B}_{\mu\nu}
+g^2c_0\epsilon^{ab3}B_{\mu\nu}
\nonumber \\
& -2ig^2(\bar{c}^ac^b-\bar{c}^cc^c\delta^{ab})\delta_{\mu\nu} ,
\label{su2u1-10}
\end{align}
where we have rescaled the gauge parameter $\alpha$ such that $g^2/\alpha\to 1/\alpha$ and completed the square for the field $A_{\mu}$. 
The off-diagonal gluon field $A_{\mu}^a$ in $S_3$ can be eliminated by the Gaussian integration,
\begin{align}
&e^{iS_0[a,B,c,\bar{c},\theta]} =
\int[dA_{\mu}^a]e^{iS_3[a,A,B,c,\bar{c};\theta]}, \\
S_0 &= -i\ln\int[dA_{\mu}^a]\exp\left\{
i\int_x 
\frac{1}{2g^2}A^{\mu}{}^aQ_{\mu\nu}^{ab}A^{\nu}{}^b
\right\} 
\nn\\
&= \frac{i}{2}\ln\det(Q_{\mu\nu}^{ab}).
\label{su2u1-11}
\end{align}
Thus we obtain the APEGT with $\theta$-term,
\begin{align} 
S_E &= S_0[a,B,c,\bar{c};\theta]+S_1[a,B;\theta] \nn\\
&+S_2[a,c,\bar{c}] .
\label{su2u1-12}
\end{align}

\subsection{Calculating the $\ln \det Q$ (Euclidean)}

The $\ln\det Q$ in 
(\ref{su2u1-11}) is divergent. To regularize it, we use 
the $\zeta$-function regularization 
\cite{Kondo97}:
\begin{align} 
\ln\det Q = -\lim_{s\to 0}\frac{d}{ds}\frac{\mu^{2s}}{\Gamma(s)}
\int_0^{\infty}dt t^{s-1}\mbox{Tr}(e^{-tQ}).
\label{lndet-2}
\end{align} 
We evaluate 
$\mbox{Tr}(e^{-tQ})$ in the Euclidean space: 
$
x_0=-i\bar{x}_0, d^4x=-id^4\bar{x} ,
$
to obtain
\begin{align} 
& \mbox{Tr}(e^{-tQ})-\mbox{Tr}(e^{-t\dd^2\delta^{ab}\delta_{\mu\nu}}) 
\nonumber\\
&= \frac{1}{16\pi^2}\int_{\bar{x}}\mbox{tr}
\left(\frac{1}{2}\tilde{Q}^2+\frac{1}{12}[D_{\mu},D_{\nu}][D_{\mu},D_{\nu}]
\right) \nn\\
& +O(t) ,
\label{lndet-4}
\end{align}
where 
\begin{align} 
\tilde{Q}_{\mu\nu}^{ab} &\equiv Q_{\mu\nu}^{ab} 
- (D_{\rho}[a]D_{\rho}[a])^{ab}\delta_{\mu\nu} 
\nonumber \\
&= -2\epsilon^{ab3}f_{\mu\nu}
+g^2c_1\epsilon^{ab3}\tilde{B}_{\mu\nu}
+g^2c_0\epsilon^{ab3}B_{\mu\nu}
\nonumber \\
& -2ig^2(\bar{c}^ac^b-\bar{c}^cc^c\delta^{ab})\delta_{\mu\nu}.
\label{lndet-5}
\end{align}
By taking into account (\ref{su2u1-6}), the trace of the second term in 
(\ref{lndet-4})  reads 
\begin{align} 
\mbox{tr}(\frac{1}{12}[D_{\mu},D_{\nu}][D_{\mu},D_{\nu}])
=-\frac{1}{3}\kappa f_{\mu\nu}f_{\mu\nu} ,
\label{lndet-6}
\end{align}
where we have introduced the  second Casimir operator $\kappa$ which is given for $G=SU(2)$ by 
\begin{align} 
\kappa\equiv C_2(G)=\epsilon^{3ab}\epsilon^{3ab}=2 .
\end{align}
The trace of the first term in (\ref{lndet-4}) reads 
\begin{align} 
& \mbox{tr}(\frac{1}{2}\tilde{Q}^2)
=\frac{1}{2}\tilde{Q}_{\mu\nu}^{ab}\tilde{Q}_{\nu\mu}^{ba}
\nonumber\\
&= 2\kappa f_{\mu\nu}f_{\mu\nu}
-\frac{1}{2}g^4\kappa B_{\mu\nu}B_{\mu\nu}
\nn\\
& -8g^4\sbra{\bar{c}^ac^b-\bar{c}^cc^c\delta^{ab}}
\sbra{\bar{c}^bc^a-\bar{c}^dc^d\delta^{ba}}
\nonumber \\
& 
-2\kappa g^2c_1 B_{\mu\nu}\tilde{f}_{\mu\nu}
-2\kappa g^2c_0 f_{\mu\nu}B_{\mu\nu}
\nn\\
& +g^4\frac{g^2\theta}{16\pi^2}\kappa \tilde{B}_{\mu\nu}B_{\mu\nu} , 
\label{lndet-7}
\end{align}
where we have used the relationship  in the Euclidean space, 
$c_1^2+c_0^2=-1$ and  $c_0c_1=g^2\theta/(16\pi^2)$. 

Hence, (\ref{lndet-2}) reads 
\begin{subequations}
\begin{align} 
& \frac{1}{2}\ln\det Q  \nn\\
&= \int_{\bar{x}} \left[
\frac{1}{4g^2}z_af_{\mu\nu}f_{\mu\nu}
+\frac{1}{4}z_bg^2B_{\mu\nu}B_{\mu\nu}
\right.\nn\\
& \left.
+\frac{1}{2}z_cB_{\mu\nu}\tilde{f}_{\mu\nu}
+\frac{1}{2}z_df_{\mu\nu}B_{\mu\nu}
+\frac{1}{2}z_eB_{\mu\nu}\tilde{B}_{\mu\nu} 
\right.\nonumber\\
& \left. +(\mbox{4-ghost terms}) 
\right.\nn\\
& \left. +(\mbox{higher derivative terms})
\right] ,
\label{lndet-8}
\end{align}
where 
\begin{align} 
z_a &= -\frac{10}{3}\kappa\frac{g^2}{16\pi^2}\ln\mu^2, \\
z_b &= \kappa\frac{g^2}{16\pi^2}\ln\mu^2, \\
z_c &= 2\kappa c_1\frac{g^2}{16\pi^2}\ln\mu^2, \\
z_d &= 2\kappa c_0\frac{g^2}{16\pi^2}\ln\mu^2, \\
z_e &= -\kappa\frac{g^4}{16\pi^2}\cdot
\frac{g^2\theta}{16\pi^2}\ln\mu^2.  
\label{lndet-9}
\end{align}
\end{subequations}
Note that $z_d, z_e \rightarrow 0$ as $\theta \rightarrow 0$. 
This result shows that the $f_{\mu\nu}f_{\mu\nu}$ and 
$B_{\mu\nu}B_{\mu\nu}$ terms receive no corrections  to one-loop order coming from the existence of  the  $\theta$ term. 

Going back to the Minkowski spacetime by taking into account 
 $\int d^4\bar{x}=i\int d^4x$ in (\ref{lndet-8}),
we obtain
\begin{align} 
& \frac{i}{2}\ln\det Q \nn\\
&= -\int_x \left[
\frac{1}{4g^2}z_af_{\mu\nu}f^{\mu\nu}
+\frac{1}{4}z_bg^2B_{\mu\nu}B^{\mu\nu}
\right.\nn\\
& \left.
+\frac{1}{2}z_cB_{\mu\nu}\tilde{f}^{\mu\nu}
+\frac{1}{2}z_df_{\mu\nu}B^{\mu\nu}
+\frac{1}{2}z_eB_{\mu\nu}\tilde{B}^{\mu\nu} 
\right.\nonumber\\
& \left. +(\mbox{4-ghost terms})
\right.\nn\\
& \left.
+(\mbox{higher derivative terms})
\right].
\label{lndet-10}
\end{align}
Thus we obtain
\begin{align} 
& S_0+S_1 \nn\\
&= \int_x \left[
-\frac{1+z_a}{4g^2}f_{\mu\nu}f^{\mu\nu}
-\frac{1+z_b}{4}g^2B_{\mu\nu}B^{\mu\nu}
\right.\nonumber\\
& \left.
-\frac{1}{2}z_cB_{\mu\nu}\tilde{f}^{\mu\nu} 
+\frac{\theta}{32\pi^2}f_{\mu\nu}\tilde{f}^{\mu\nu} 
+\frac{1}{2}z_df_{\mu\nu}B^{\mu\nu}
\right.\nonumber\\
& \left.
+\frac{1}{2}z_eB_{\mu\nu}\tilde{B}^{\mu\nu}
+ (\mbox{4-ghost terms})
\right.\nonumber\\
& \left.
+(\mbox{higher derivative terms})
\right] .
\label{lndet-11}
\end{align}

\underline{Integrating the off-diagonal ghost field $c^a$, $\bar{c}^a$}

Neglecting the 4-ghost term and the higher derivative terms in 
(\ref{lndet-11}), the integration over $c^a$ and $\bar{c}^a$ in (\ref{su2u1-8}) reduces to the Gaussian type and can be performed easily \cite{Kondo97}. 
The result to one-loop order reads 
\begin{subequations}
\begin{align} 
iS_c &= \ln\int [d\bar{c}][dc]\exp
\mbra{i\int_x i\bar{c}^aD^{\mu ac}[a]D_{\mu}^{cb}[a]c^b} \nonumber\\
&= i\int_x \frac{1}{4g^2}z'_af_{\mu\nu}f^{\mu\nu}+\cdot\cdot\cdot ,
\label{lndet-13}
\end{align}
where
\begin{align} 
z'_a &= \frac{1}{3}\kappa\frac{g^2}{16\pi^2}\ln\mu^2.
\label{lndet-14}
\end{align}
\end{subequations}

\section{APEGT with dyon}

The U(1) antisymmetric tensor field $B_{\mu\nu}$ has the Hodge-de Rham decomposition,
\begin{subequations}
\begin{align} 
B_{\mu\nu} &= b_{\mu\nu}+\tilde{\chi}_{\mu\nu}, \\
b_{\mu\nu} &= \dd_{\mu}b_{\nu}-\dd_{\nu}b_{\mu}, \\
\tilde{\chi}_{\mu\nu} &= \frac{1}{2}\epsilon_{\mu\nu\alpha\beta}
(\dd^{\alpha}\chi^{\beta}-\dd^{\beta}\chi^{\alpha}) .
\label{apegt-1}
\end{align}
\end{subequations}
Then the integration measure for $B_{\mu\nu}$ is replaced by the measure for 
$b_{\mu}$ and $\chi_{\mu}$, 
\begin{align} 
[dB_{\mu\nu}]=[db_{\mu}][d\chi_{\mu}]\delta(F[b])\delta(F[\chi]) ,
\label{apegt-2}
\end{align}
where $F[b]$ and $F[\chi]$ are gauge fixing conditions for the gauge symmetries, 
\begin{align} 
b_{\mu}(x) & \rightarrow b_{\mu}(x)-\dd_{\mu}\theta(x) ,
\nonumber\\
\chi_{\mu}(x)  & \rightarrow  \chi_{\mu}(x)-\dd_{\mu}\varphi(x) .
\label{apegt-3}
\end{align}
Thus, 
(\ref{lndet-11}) reads
\begin{align} 
& S_0+S_1 \nn\\
&= \int_x \left[
-\frac{1+z_a}{4g^2}f_{\mu\nu}f^{\mu\nu}
-\frac{1+z_b}{4}g^2(b_{\mu\nu}b^{\mu\nu}
\right.\nn\\
& \left.
+\tilde{\chi}_{\mu\nu}\tilde{\chi}^{\mu\nu})
-\frac{1}{2}z_cb_{\mu\nu}\tilde{f}^{\mu\nu} 
-\frac{1}{2}z_c\tilde{\chi}_{\mu\nu}\tilde{f}^{\mu\nu} 
\right.\nonumber\\
& \left.
+\frac{1}{4}\frac{\theta}{8\pi^2}f_{\mu\nu}\tilde{f}^{\mu\nu} 
+\frac{1}{2}z_df_{\mu\nu}b^{\mu\nu}
+\frac{1}{2}z_df_{\mu\nu}\tilde{\chi}^{\mu\nu}
\right.\nonumber\\
& \left.
+\frac{1}{2}z_eb_{\mu\nu}\tilde{b}^{\mu\nu}
+z_eb_{\mu\nu}\chi^{\mu\nu}
\right] .
\label{apegt-4a}
\end{align}
Here integrating out the $\chi_{\mu}$ field yields the corrections to 
$ff$-, $f\tilde{f}$-, $bb$-terms.  However, they do not affect the one-loop results.  Therefore, the effective action to one-loop order 
is obtained by taking into account the contribution (\ref{lndet-14}) from the ghost as
\begin{subequations}
\begin{align} 
S_E &= \int_x \left[
-\frac{1+z_a-z'_a}{4g^2}f_{\mu\nu}f^{\mu\nu}
+\frac{1}{4}\frac{\theta}{8\pi^2}f_{\mu\nu}\tilde{f}^{\mu\nu} 
\right.\nn\\
& \left.
-\frac{1+z_b}{4}g^2b_{\mu\nu}b^{\mu\nu}
-\frac{1}{2}z_cb_{\mu\nu}\tilde{f}^{\mu\nu} 
\right.\nonumber\\
& \left.
+\frac{1}{2}z_df_{\mu\nu}b^{\mu\nu}
+\frac{1}{2}z_eb_{\mu\nu}\tilde{b}^{\mu\nu}
\right] .
\label{apegt-4}
\end{align}
Here the last two terms in (\ref{apegt-4}) are cast into 
\begin{align} 
f_{\mu\nu}b^{\mu\nu} &= -2\dd_{\nu}b_{\mu}f^{\mu\nu} 
=2b_{\mu}\dd_{\nu}f^{\mu\nu} 
\nn \\
& +(\mbox{surface term}), \\
b_{\mu\nu}\tilde{b}^{\mu\nu} 
&= 2\epsilon^{\mu\nu\rho\sigma}\dd_{\mu}b_{\nu}\dd_{\rho}b_{\sigma}
=-2\epsilon^{\mu\nu\rho\sigma}b_{\nu}\dd_{\mu}\dd_{\rho}b_{\sigma}
\nn \\
&+(\mbox{surface term}).
\label{apegt-5}
\end{align}
\end{subequations}
and they are neglected, provided that $\dd_{\nu}f^{\mu\nu}=J^{\mu}=0$ and $b_{\mu}$ is regular. 

Defining the  magnetic current $k_{\mu}$ by 
\begin{align} 
k^{\mu}\equiv \dd_{\nu}\tilde{f}^{\mu\nu}, \quad 
\tilde{f}^{\mu\nu}=\frac{1}{2}\epsilon^{\mu\nu\rho\sigma}f_{\rho\sigma} ,
\label{apegt-6}
\end{align}
we obtain the APEGT including the magnetic current $k_{\mu}$ from (\ref{apegt-4}) 
\begin{align} 
S_E &= \int_x \left[
-\frac{1}{4}Z_a^{-1}f_{\mu\nu}f^{\mu\nu}
+\frac{g\theta}{16\pi^2}a_{\mu}k^{\mu}
\right. \nn \\
&\left.
-\frac{1}{4}Z_b^{-1}b_{\mu\nu}b^{\mu\nu}
-\frac{1}{2g}z_cb_{\mu}k^{\mu}
\right] ,
\label{apegt-7}
\end{align}
where we have defined 
\begin{align} 
Z_a &\equiv (1+z_a-z'_a)^{-1}=1-z_a+z'_a, 
\nonumber\\
Z_b &\equiv (1+z_b)^{-1}=1-z_b ,
\label{apegt-8}
\end{align}
and rescaled $a_{\mu}/g\to a_{\mu}$ and $gb_{\mu}\to b_{\mu}$.

We define the wave function renormalization for 
$a_{\mu}$ and $b_{\mu}$ by 
\begin{align} 
a_{\mu}^R \equiv Z_a^{-1/2}a_{\mu}, 
\quad
b_{\mu}^R \equiv Z_b^{-1/2}b_{\mu}.
\label{apegt-9}
\end{align}
Then (\ref{apegt-7}) is cast into the renormalized form,
\begin{align} 
S_E &= \int_x \left[
-\frac{1}{4}f_{\mu\nu}^Rf^{R\mu\nu}
-\frac{1}{4}b_{\mu\nu}^Rb^{R\mu\nu}
\right.\nn\\
&\left.
+\frac{g^R\theta^R}{2\pi}a_{\mu}^Rk^{R\mu}
-\frac{4\pi}{g^R}b_{\mu}^Rk^{R\mu}
\right] ,
\label{apegt-10}
\end{align}
where we have defined the renormalized quantities,
\begin{subequations}
\begin{align} 
g^R &\equiv Z_a^{1/2}g, \\
\theta^R &\equiv Z_a^{-1/2}Z_c^{-1}Z_b^{-1/2}\theta, \\
k^{R\mu} &\equiv Z_a^{1/2}Z_cZ_b^{1/2}k^{\mu} ,
\label{apegt-11}
\end{align}
\end{subequations}
and rescaled $k^R_{\mu}\to 8\pi k^R_{\mu}$. 
We find that $k_{\mu}$ has the renormalization factor $Z_a$ 
due to the existence of the $\theta$ term.
The existence of the third term, the cross term of the magnetic current $k_\mu$ with the electric field $a_\mu$,  in (\ref{apegt-10}) indicates that 
{\it the monopole current $k_{\mu}$  acquires the electric charge and the magnetic monopole is changed to the dyon due to the existence of the $\theta$ term} in agreement with the Witten effect  
 \cite{Witten:1979ey2}.
\footnote{
The dyon treated in this paper goes to the usual magnetic monopole in the vanishing $\theta$ angle limit, $\theta \rightarrow 0$.  Therefore, it is different from the usual (old-fashioned) dyon \cite{JZ75,BPS75} which is the 
magnetic monopole having a non-vanishing electric charge even for $\theta=0$.
}
In what follows, we omit the index $R$ of the field. 

We observe that the Lagrangian
\begin{align} 
\mathcal{L}_0 &= -\frac{1}{4}f_{\mu\nu}f^{\mu\nu}
-\frac{1}{4}b_{\mu\nu}b^{\mu\nu}
\label{apegt-12}
\end{align}
is invariant under the linear transformation for $a_{\mu}$ and $b_{\mu}$ 
with an arbitrary constant $v$:
\begin{align} 
\vect{a_{\mu}}{b_{\mu}} =
\mtx{\cos v}{\sin v}{-\sin v}{\cos v}
\vect{a'_{\mu}}{b'_{\mu}} .
\label{apegt-13}
\end{align}
By choosing  
\begin{align} 
v=\arctan \sbra{-\frac{g^2\theta}{8\pi^2}} ,
\label{apegt-14}
\end{align}
we can eliminate the cross term in  
(\ref{apegt-10}) which is transformed into 
\begin{align} 
S_E = \int_x \left[
-\frac{1}{4}f'_{\mu\nu}f'{}^{\mu\nu}
-\frac{1}{4}b'_{\mu\nu}b'{}^{\mu\nu}
-g_m[\theta]b'_{\mu}k'{}^{\mu}
\right] ,
\label{apegt-15}
\end{align}
where 
\begin{align} 
g_m[\theta]  := g |\tau|  = \sqrt{g_m^2 + q_m^2} ,
\nonumber\\
 g_m \equiv \frac{4\pi}{g},
\quad q_m \equiv \frac{g\theta}{2\pi} .
\label{apegt-16}
\end{align}
Finally, by integrating out the field $b_{\mu}'$, we obtain (omitting the prime in what follows)
\begin{align} 
S_E =  \int_x \left[
-\frac{1}{4}f_{\mu\nu}f^{\mu\nu}
+\frac{1}{2}g_m^2[\theta]k^{\mu}D_{\mu\nu}k^{\nu}
\right] ,
\label{apegt-17}
\end{align}
where the kernel $D_{\mu\nu}$ stands for the massless vector propagator obtained after introducing the gauge fixing term for $b_{\mu}'$, e.g., 
$D_{\mu\nu}=(1/\partial^2)(\delta_{\mu\nu} - \partial_\mu \partial_\nu/\partial^2)$ in the Landau gauge. 
It is remarkable that the effect of the $\theta$ angle is combined into a compact form written in terms of the complex coupling constant $\tau$ even after the Abelian projection, since
$g_m[\theta]=g|\tau|$. 

This result should be compared with the effective theory (3.4) of \cite{EI82} written in  terms of the electric current $j_\mu$ and the magnetic current $k_\mu$.  Indeed, if the electric current $j_\mu$ is eliminated in (3.4) of \cite{EI82}, the resulting theory agrees with our result (\ref{apegt-17}). 
However, it was assumed in \cite{EI82} that the Yang-Mills theory is approximated in terms of Abelian fields with the $\theta$ angle at a long-distance scale $R$. 

\section{Topological susceptibility and Witten-Veneziano formula}

 Now we argue that {\it the dyon configuration is the most relevant one
 for solving the U(1) problem} in SU(2) QCD by evaluating the
 topological susceptibility from the dyon 
 configuration appearing in the APEGT with $\theta$-term.  

Integrating out $a_{\mu}$ in (\ref{apegt-17}), we obtain the {\it effective dyon action}
\begin{align} 
S_E = \int_x \left[
\frac{1}{2}\mbra{\sbra{\frac{4\pi}{g}}^2
+\sbra{\frac{g\theta}{2\pi}}^2}k^{\mu}D_{\mu\nu}k^{\nu}
\right] .
\label{WV-1}
\end{align}
To estimate the numerical value of the topological susceptibility, we consider the lattice regularized version of (\ref{WV-1}), 
\begin{align} 
S_E &= \sum_{x,y}\sbra{\bar{\beta}+\frac{\theta^2}{\bar{\beta}}}
k^{\mu}(x)D_{\mu\nu}(x-y)k^{\nu}(y), 
\nonumber\\
\bar{\beta} &\equiv \frac{1}{2}\sbra{\frac{4\pi}{g}}^2.
\label{WV-2}
\end{align}
The part of the self-mass term $k^{\mu}(x) k^{\mu}(x) $ is extracted from 
(\ref{WV-2}) as%
\footnote{
According to the analysis of the monopole action by the inverse Monte-Carlo simulation,  the self-mass term of the monopole current is dominant in the low-energy region, e.g., $G_2/G_1\simeq 0.33$ at the scale $1.7 \rm{fm}$ where 
 $G_1$ and $G_2$ are respectively the self-coupling and the nearest-neighbor coupling of the monopole current \cite{Kato:2000}.
}
\begin{align} 
S_E \simeq \sbra{\bar{\beta}+\frac{\theta^2}{\bar{\beta}}}D(0)
\sum_xk^{\mu}(x)k^{\mu}(x) ,
\label{WV-3}
\end{align}
where $D(0)<\infty$ on a lattice.
Furthermore, as the monopole configuration subject to%
\footnote{
The monopole current is integer-valued on the lattice, if the construction due to DeGrand and Toussaint \cite{DT80} is used.
}
 $|k_{\mu}(x)|=1$ is dominant in the low-energy region \cite{SS91}, the energy density $e_\theta$ is written as 
\begin{align} 
e_\theta = S_E/V \simeq \sbra{\bar{\beta}+\frac{\theta^2}{\bar{\beta}}}D(0) .
\label{WV-4}
\end{align}
Therefore, the topological susceptibility $\chi_E$ is calculated:%
\footnote{
Here $\theta$ should be understood as the renormalized variable $\theta_R$.
}
\begin{align} 
\chi_E \equiv \sbra{\frac{d^2 e_\theta}{d\theta^2}}_{\theta=0}
\simeq \frac{2}{\bar{\beta}}D(0) .
\label{WV-5}
\end{align}
The result of Chernodub et al.\cite{Kato:2000} show 
$\bar{\beta}D(0)\equiv G_1 =0.059$ and $\bar{\beta}=2.49$
at the physical scale 
$b=3.8\sigma_{phys}^{-1/2}$. 
(Note that 
$b=1\sigma_{phys}^{-1/2}$ corresponds to $1.7 \rm{fm}$, provided that the string tension $\sigma_{phys} \cong (440\rm{MeV})^2$ in SU(2) QCD.
)
 By substituting these values into (\ref{WV-5}), the topological susceptibility is determined as%
\footnote{
In this section, we have used a quantum perfect monopole action to evaluate 
the topological susceptibility $\chi_E$, that is an action on the renormalized
trajectory on which one can take the continuum limit. Therefore, our prediction
agree with those of the continuum independently whether the lattice is fine or 
coarse. See \cite{FKS:2000} for the detail of the quantum perfect action.
}
\begin{align} 
\chi_E^{1/4}/\sigma_{phys}^{1/2}=0.371 ,
\label{WV-6}
\end{align}
in units of the string tension $\sigma_{phys}$. 
Remarkably, this estimate reproduces $76\%$ of the full result 
\begin{align}
  \chi^{1/4}/\sigma_{phys}^{1/2}=0.486\pm 0.010 ,
\label{Teper}
\end{align}
obtained by  Teper \cite{Teper98} in the simulation of SU(2)QCD. 
Moreover, our result is also consistent with those of 
Bornyakov and Schierholz~\cite{BS96} and 
Sasaki and Miyamura~\cite{SM99} 
where the Abelian dominance for $\chi$ was reported based on the numerical simulations.

 For large $N_c$ in the $SU(N_c)$ QCD,  
Witten \cite{Witten:1979ey1} has shown by taking into account the next-to-leading order of $1/N_c$ expansion that the U(1) chiral symmetry is broken due to the axial-vector anomaly and hence the NG boson $\eta'$ can acquire the non-zero mass of ${\cal O}(N_c^{-1})$ even in the chiral limit.
Moreover, Witten has derived the mass formula for $\eta'$, the so-called the  Witten-Veneziano formula\cite{Veneziano79}: 
\begin{align} 
m_{\eta'}^2=\frac{4N_f}{f_{\pi}^2}
\sbra{\frac{d^2E_{\theta}}{d\theta^2}}_{\theta=0}^{\mbox{no-quarks}} ,
\label{WV-7}
\end{align}
where $E_{\theta}$ is the vacuum energy density of the gluon field. 
Substituting the numerical values, 
$m_{\eta'}\simeq 1\rm{GeV}$ and $f_{\pi}\simeq 0.1\rm{GeV}$
into (\ref{WV-7}), the topological susceptibility is estimated as (for $N_f=3$)
\begin{align} 
\chi &\equiv
\sbra{\frac{d^2E_{\theta}}{d\theta^2}}_{\theta=0}^{\mbox{no-quarks}} \nn \\
&= \frac{1}{12}(0.1\rm{GeV})^2(1\rm{GeV})^2  \nn\\
&\simeq 8\times 10^{-4}(\rm{GeV})^4 ,
 \nonumber\\
\to \chi^{1/4} &\simeq  150 \sim 200\rm{MeV} .
\label{WV-7a}
\end{align}

This result for large $N_c$ of $SU(N_c)$ case is consistent with the SU(2) result (\ref{Teper}). 
In order for this formula to be meaningful, the vacuum energy density $E_{\theta}$ must depend on the $\theta$ angle. 

Although we have restricted to the SU(2) gauge group in this analysis, these results suggest that the large $N_c$ analysis in the next-to-leading order gives fairly good estimate also for $N_c=2$ in the U(1) problem. 
In fact,  this claim is also confirmed by the numerical simulations on the lattice for various $N_c$ of SU($N_c$) Yang-Mills theory, see Teper \cite{Teper98}.

Thus we conclude  that {\it the dyon, i.e., magnetic monopole with the electric charge proportional to the vacuum angle $\theta$, gives dominant contribution to the topological susceptibility}.%
\footnote{
Note that the above estimate of the topological susceptibility was obtained based on the self-mass term alone.  It is expected therefore that the inclusion of the remaining terms such as the nonlocal interaction terms reproduce the whole topological susceptibility of the original SU(2) gluodynamics. 
}

\section{Conclusion and discussion}

In this paper, we have argued an interesting possibility that the U(1) problem is solved by the dyonic configuration appearing in the APEGT with a vacuum angle $\theta$.
For this purpose, we started with the Yang-Mills theory with a vacuum angle $\theta$ in the MAG.  
We have separated the Abelian component by exploiting the idea of the Abelian projection and then integrated out all the off-diagonal components except for the diagonal ones.  
Applying a duality transformation to the resulting theory, we have obtained an effective theory written in terms of the dyon degrees of freedom, called the APEGT with $\theta$-term. 
By making use of the classical part of the dyon action, we have estimated the topological susceptibility.  The obtained value agrees with the numerical result obtained by the recent lattice gauge theory. 
Thus we have shown that the dyon configuration generated by the vacuum angle $\theta$ gives a dominant contribution to the topological susceptibility and resolves the U(1) problem.

In this paper, we have treated only the SU(2) case in detail. 
In order to confirm the consistency of our claim with the large $N_c$ result,  
it is desirable to extend our method to SU(3) case.  This will be reported in a subsequent paper \cite{KK04}.

In the derivation above, there is a subtle point to be mentioned. 
To estimate the topological susceptibility, we have translated the continuum result (\ref{WV-1}) to the lattice one (\ref{WV-2}), and used a fact that $D(0)$ is finite on the lattice. 
However, the kernel $D_{\mu\nu}$ in the continuum (\ref{WV-1}) does not have the 
contact interaction, and hence the self-mass term of the monopole current does not exist in the rigorous sense within this derivation.  
Therefore, it should be understood that we have introduced a cutoff in (\ref{WV-3}) to regularize $D_{\mu\nu}(x-y)$ so that $D(0)<\infty$, just as Ezawa and Iwazaki \cite{EI82} replaced $D_{\mu\nu}(x-y)$ by the massless propagator with the momentum cutoff $R^{-1}$.

Such a physical cutoff naturally appears if the off-diagonal gluons acquire their mass. 
In fact, the numerical simulations on the lattice have confirmed the non-zero mass for the off-diagonal gluons \cite{AS99}. 
Some analytical studies in this direction exist too, see \cite{Schaden99}.
It is possible to show that the self-interaction term between monopole currents and the derivative term appear as a consequence of mass generation \cite{Kondo01}.  
This point deserves further studies in connection with the U(1) problem \cite{KK04}.

We comment on the role of instantons to the U(1) problem. 
In this paper we have stressed that the magnetic monopole promoted to the dyon due to the vacuum angle $\theta$ resolves the U(1) problem. 
Our results suggest the compatibility between the dyon configuration with the large $N_c$ expansion, although it is not definitive. 
Although instantons and magnetic monopole are originated from different non-trivial homotopy groups,
$\pi_3(SU(N_c))=\mathbb{Z}$ and $\pi_2(SU(N_c)/U(1)^{N_c-1})=\mathbb{Z}^{N_c-1}$ respectively \cite{Actor79},  
the strong correlation between instantons and magnetic monopoles have been reported recently for $N_c=2$, see e.g., \cite{BOT97} for analytical works and \cite{STSM96} for numerical works.
Moreover, there is a vast and consistent literature concerning the lattice determination of the pure-gauge topological susceptibility on the lattice, in which the role of instantons for $N_c=2$ and $N_c=3$ has been well proved and tested by means of the so-called {\it cooling method}, 
see e.g., \cite{Lattice-instanton-su2} for SU(2), 
\cite{Lattice-instanton-su3} for SU(3)
and \cite{SS98} for a review.
The large-$N_c$ behaviour of instantons is not so clear and far from definite, as discussed already in the classic papers \cite{BZ80}. 
Thus we arrive at a viewpoint that the magnetic monopole and the dyon should be treated on an equal footing with the instanton as non-perturbative topological configurations to be taken account of in solving the U(1) problem.

In this paper, we have discussed the U(1) problem only within the framework of the pure Yang-Mills theory with the vacuum angle $\theta$. 
Suppose that the quark degrees of freedom are introduced into the consideration.  Then it is interesting to study the relationship of our results with the 
Atiyah--Singer index theorem \cite{AS68} (see also chapter 11 of \cite{Rajaraman82}), 
the fermionic zero modes (low-lying eigenvalue of the Dirac operator), 
and Banks--Casher formula \cite{BC80}.
Such investigations have already been done on a lattice by numerical simulations, e.g., see \cite{IIY87} in the instanton background and \cite{SM98} in the monopole background. 
These results also demonstrate the strong correlation between instantons and magnetic monopoles.
The investigation from the analytical side is a future task.

\section*{Acknowledgments}
This work is supported by 
Grant-in-Aid for Scientific Research (C)14540243 from Japan Society for the Promotion of Science (JSPS), 
and in part by Grant-in-Aid for Scientific Research on Priority Areas (B)13135203 from
the Ministry of Education, Culture, Sports, Science and Technology (MEXT).

\baselineskip 12pt
\def\vol#1#2#3{{\bf {#1}} ({#2}) {#3}}
\def\NP{Nucl.~Phys. }
\def\PL{Phys.~Lett. }
\def\PR{Phys.~Rev. }
\def\PRL{Phys.~Rev.~Lett. }
\def\PTP{Prog.~Theor.~Phys. }
\def\MPL{Mod.~Phys.~Lett. }
\def\IJMP{Int.~J.~Mod.~Phys. }
\def\JETP{Sov.~Phys.~JETP }
\def\JP{J.~Phys. }
\def\NPPS#1#2#3{%
\NP (Proc.~Suppl. {\bf {#1}}) ({#2}) {#3}}
\def\PhR{{\it Phys}.~{\it Report}. }
\def\NP{{\it Nucl}.~{\it Phys}. }
\def\PL{{\it Phys}.~{\it Lett}. }
\def\PR{{\it Phys}.~{\it Rev}. }
\def\PRL{{\it Phys}.~{\it Rev}.~{\it Lett}. }
\def\PTP{{\it Prog}.~{\it Theor}.~{\it Phys}. }
\def\MPL{{\it Mod}.~{\it Phys}.~{\it Lett}. }
\def\IJMP{{\it Int}.~{\it J}.~{\it Mod}.~{\it Phys}. }
\def\JETP{{\it Sov}.~{\it Phys}.~{\it JETP} }
\def\JP{{\it J}.~{\it Phys}. }
\def\NPPS#1#2#3{%
\NP {\it Proc}.~{\it Suppl}. {\bf {#1}} ({#2}) {#3}}
\def\th#1{hep-th/{#1}}
\def\ph#1{hep-ph/{#1}}



\begin{thebibliography}{99}
\bibitem{Weinberg75}
S. Weinberg, 
Phys. Rev. D11, 3583-3593 (1975).


\bibitem{tHooft76}
G. 't Hooft,
Phys. Rev. Lett. 37, 8-11 (1976).
\\
G. 't Hooft,
Phys. Rept. 142, 357-387 (1986). 


\bibitem{instantonGroup}
R. Jackiw and C. Rebbi,
Phys. Rev. Lett. 37, 172 (1976).
\\
C.G. Callan, R. Dashen and D.J. Gross,
Phys. Lett. B 63, 334 (1976).

\bibitem{Crewther77}
 R.J. Crewther,
Phys. Lett. B70, 349--354 (1977).


\bibitem{Witten:1979ey1} 
E. Witten,
Nucl. Phys. B 156, 269--283 (1979).

\bibitem{Veneziano79}
G. Veneziano, 
Nucl. Phys. B 159, 213--224 (1979).


\bibitem{U1}
  P. Di Vecchia,
Phys. Lett. 85B, 357--360 (1979).
\\
 C. Rosenzweig, J. Schechter and G. Trahern,
Phys. Rev. D21, 3388--3392 (1980).
\\
 P. Di Vecchia and G. Veneziano,
Nucl. Phys. B 171, 253--272 (1980).
\\
 P. Nath and R. Arnowitt
Phys. Rev. D23, 473--476 (1981). 
\\
 K. Kawarabayashi and N. Ohta,
Nucl. Phys. B 175, 477--492 (1980).
\\
K. Kawarabayashi and N. Ohta,
Prog. Theor. Phys. 66, 1789--1802 (1981).
\\
  T. Kugo,
Nucl. Phys. B 155, 368--380 (1979).
\\
H. Hata, T. Kugo and N. Ohta,
Nucl. Phys. B 178, 527--544 (1981).


\bibitem{tHooft81} 
  G. 't Hooft,
  Nucl.Phys. B 190 [FS3], 455-478 (1981).


\bibitem{EI82}
  Z.F. Ezawa and A. Iwazaki,
Phys. Rev. D 26, 631--647 (1982).


\bibitem{Kondo97}  
  K.-I. Kondo,
  Phys. Rev. D 57, 7467--7487 (1998).
\\
K.-I. Kondo and T. Shinohara,
Prog. Theor. Phys. 105, 649--665 (2001).
\\
T. Shinohara,
Mod. Phys. Lett. A18, 1398--1412 (2003).


\bibitem{AS99}
  K. Amemiya and H. Suganuma,
Phys. Rev. D 60, 114509 (1999).
\\
  V.G. Bornyakov, M.N. Chernodub, F.V. Gubarev, S.M. Morozov and M.I. Polikarpov, 
 Phys. Lett. B559, 214--222 (2003).


\bibitem{Harvey96}
J.A. Harvey,
hep-th/9603086.


\bibitem{Witten:1979ey2} 
E.~Witten,
Phys. Lett.  86B, 283--287 (1979).


\bibitem{JZ75}
 B. Julia and A. Zee,
Phys. Rev. D11, 2227--2232 (1975).


\bibitem{BPS75}
M.K. Prasad and C. Sommerfield,
Phys. Rev. Lett. 35, 760--762 (1975).
\\
E.B. Bogomol'nyi, 
Sov. J. Nucl. Phys. 24, 449 (1976).


\bibitem{Teper98}
M.Teper,
hep-th/9812187.


\bibitem{SS91} 
J.~Smit and A.J.van der Sijs,
Nucl. Phys. B335, 603--648 (1991).

\bibitem{DT80}
T.A. DeGrand and D. Toussaint, 
Phys. Rev. D22, 2478-2489 (1980).

\bibitem{Kato:2000} 
M.N. Chernodub, S. Fujimoto, S. Kato, M. Murata, M.I. Polikarpov and T. Suzuki,
Phys. Rev. D62, 094506 (2000). 


\bibitem{FKS:2000}
S.~Fujimoto, S.~Kato, T.~Suzuki, 
Phys.Lett. B476, 437-447  (2000).


\bibitem{BS96} 
V. Bornyakov and G. Schierholz, 
Phys. Lett. B384, 190--196 (1996).


\bibitem{SM99} 
S. Sasaki and O. Miyamura,
Phys. Rev. D59, 094507 (1999). 


\bibitem{Schaden99} 
  M. Schaden,
  hep-th/9909011.
\\
  K.-I. Kondo and T. Shinohara,
  Phys. Lett. B 491, 263--274 (2000).
\\
D. Dudal and H. Verschelde,
J. Phys. A 36, 8507--8516 (2003).
\\
  K.-I. Kondo,
  Phys. Lett. B 514, 335--345 (2001).  
\\
  K.-I. Kondo, T. Murakami, T. Shinohara and T. Imai,
  Phys. Rev. D 65, 085034 (2002).
\\
  K.-I. Kondo, 
Phys. Lett. B 572, 210--215 (2003). 
\\
K.-I. Kondo,
hep-th/0404252, 
Phys. Lett. B, to appear.
\\
D. Dudal, J.A. Gracey, V.E.R. Lemes, M.S. Sarandy, R.F. Sobreiro, S.P. Sorella and H. Verschelde, 
hep-th/0406132. 
\\
A.A. Slavnov, 
hep-th/0407194.


\bibitem{Kondo01}
 K.-I. Kondo,
  hep-th/0009152.
\\
  F. Freire,
Phys. Lett. B 526, 405--412 (2002).


\bibitem{KK04} 
S.~Kato and K.-I.~Kondo,
in preparation.


\bibitem{Actor79}
R. Actor, 
Rev. Mod. Phys. 51, 461--525 (1979).


\bibitem{BOT97}
R.C. Brower, K.N. Orginos and C.-I. Tan,
Phys. Rev. D 55, 6313-6326 (1997).


\bibitem{STSM96}
H. Suganuma, A. Tanaka, S. Sasaki and O. Miyamura,
Nucl. Phys. B (Proc. Suppl.) 47, 302--305 (1996).
\\
A. Hart and M. Teper,
[hep-lat/9511016] 
Phys. Lett. B 371, 261--269 (1996).
\\
S. Thurner, M. Feurstein, H. Markum, W. Sakuler,
Phys.Rev.D54, 3457--3464 (1996).
\\
S. Thurner, M.C. Feurstein and H. Markum,
Phys. Rev. D56, 4039--4042 (1997).
\\
M. Feurstein, H. Markum, S. Thurner,
Phys.Lett. B396, 203--209 (1997). 


\bibitem{Lattice-instanton-su2}
B. Alles, M. D'Elia, A. Di Giacomo and R. Kirchner,
[hep-lat/9709074], 
Nucl. Phys. Proc. Suppl. 63, 510-512 (1998). 
\\
P. de Forcrand, M.G. Perez, I.-O. Stamatescu,
[hep-lat/9701012], 
Nucl. Phys. B499, 409-449 (1997). 
\\
B. Alles, M. D'Elia and A. Di Giacomo,
[hep-lat/9706016],  
Phys. Lett. B412, 119-124 (1997). 
\\
C. Michael and P.S. Spencer,
[hep-lat/9503018],
Phys. Rev. D52, 4691-4699 (1995). 
\\
B. Alles, M. Campostrini, A. Di Giacomo, Y. Gunduc and E. Vicari,
[hep-lat/9302004], 
Phys. Rev. D48, 2284-2289 (1993). 
\\
B. Alles, A. Di Giacomo and M. Giannetti,
Phys. Lett. B249, 490-494 (1990).


\bibitem{Lattice-instanton-su3}
B. Alles, M. D'Elia and A. Di Giacomo, 
[hep-lat/0411035]
Phys. Rev. D71, 034503 (2005). 
\\
L. Del Debbio, L. Giusti and C. Pica, 
[hep-th/0407052], 
Phys. Rev. Lett. 94, 032003 (2005). 
hep-lat/0409100. 
\\
B. Alles, M. D'Elia and A. Di Giacomo, 
[hep-lat/9605013],  
Nucl. Phys. B494, 281-292 (1997), Erratum-ibid. B679, 397-399 (2004).
\\
P. de Forcrand, M.G. Perez and I.-O. Stamatescu, 
[hep-lat/9608032], 
Nucl. Phys. Proc. Suppl. 53, 557-559 (1997).
\\
M.C. Chu, J.M. Grandy, S. Huang and  J.W. Negele,
[hep-lat/9312071],
Phys. Rev. D49, 6039-6050 (1994). 


\bibitem{SS98}
T. Schafer and E.V. Shuryak,
[hep-ph/9610451], 
Rev. Mod. Phys. 70, 323-426 (1998). 

\bibitem{BZ80}
  W.A. Bardeen and V.I. Zakharov, 
  Phys. Lett. 91B, 111 (1980). 
\\
  E.V. Shuryak, 
  Nucl. Phys. B 203, 93 (1982).
\\
  D.I. Dyakonov and V.Yu. Petrov, 
  Nucl. Phys. B 245, 259 (1984).




\bibitem{BC80}
T. Banks and A. Casher,  
Nucl.Phys. B169, 103 (1980). 


\bibitem{AS68}
M.F. Atiyah and I.M. Singer,
Ann. Math. 87, 484--530 (1968).


\bibitem{Rajaraman82}
R.Rajaraman,
Solitons and instantons 
(North--Holland, Amsterdam, 1982).


\bibitem{IIY87}
S. Itoh, Y. Iwasaki and T. Yoshie,
Phys. Rev. D 36, 527--545 (1987).


\bibitem{SM98}
S. Sasaki and O. Miyamura,
Phys. Lett. B 443, 331--337 (1998).











\end{thebibliography}
\end{document}